\documentclass[12pt,a4paper]{article}

\usepackage{ifthen} 
\newboolean{pdflatex}
\setboolean{pdflatex}{true} 

\newboolean{articletitles}
\setboolean{articletitles}{true} 

\newboolean{uprightparticles}
\setboolean{uprightparticles}{false} 

\newboolean{inbibliography}
\setboolean{inbibliography}{false} 

\usepackage{xcolor,colortbl}

\usepackage{microtype}
\usepackage{lineno}  
\usepackage{xspace} 
\usepackage{caption} 

\usepackage{graphicx}  
\usepackage{color}
\usepackage{colortbl}
\graphicspath{{./figs/}} 

\usepackage{amsmath} 
\usepackage{amssymb}
\usepackage{amsfonts}
\usepackage{upgreek} 

\newcommand*\patchAmsMathEnvironmentForLineno[1]{%
\expandafter\let\csname old#1\expandafter\endcsname\csname #1\endcsname
\expandafter\let\csname oldend#1\expandafter\endcsname\csname
end#1\endcsname
 \renewenvironment{#1}%
   {\linenomath\csname old#1\endcsname}%
   {\csname oldend#1\endcsname\endlinenomath}%
}
\newcommand*\patchBothAmsMathEnvironmentsForLineno[1]{%
  \patchAmsMathEnvironmentForLineno{#1}%
  \patchAmsMathEnvironmentForLineno{#1*}%
}
\AtBeginDocument{%
\patchBothAmsMathEnvironmentsForLineno{equation}%
\patchBothAmsMathEnvironmentsForLineno{align}%
\patchBothAmsMathEnvironmentsForLineno{flalign}%
\patchBothAmsMathEnvironmentsForLineno{alignat}%
\patchBothAmsMathEnvironmentsForLineno{gather}%
\patchBothAmsMathEnvironmentsForLineno{multline}%
\patchBothAmsMathEnvironmentsForLineno{eqnarray}%
}

\usepackage{hyperref}    
\usepackage[all]{hypcap} 

\def\lhcb {\mbox{LHCb}\xspace}

\def\MagUp {\mbox{\em Mag\kern -0.05em Up}\xspace}

\ifthenelse{\boolean{uprightparticles}}%
{

 \def\Ppi         {\ensuremath{\uppi}\xspace}                 
                  
 \def\Prho        {\ensuremath{\uprho}\xspace}

 \def\PDelta      {\ensuremath{\Delta}\xspace}                 
 \def\PXi      {\ensuremath{\Xi}\xspace}                 
 \def\PLambda      {\ensuremath{\Lambda}\xspace}                 
 \def\PSigma      {\ensuremath{\Sigma}\xspace}                 
 \def\POmega      {\ensuremath{\Omega}\xspace}                 
 \def\PUpsilon      {\ensuremath{\Upsilon}\xspace}                 
 
 \def\PB      {\ensuremath{\mathrm{B}}\xspace}                 
                  
 \def\PD      {\ensuremath{\mathrm{D}}\xspace}

 \def\PK      {\ensuremath{\mathrm{K}}\xspace}

 \def\Pi      {\ensuremath{\mathrm{i}}\xspace}

}
{

 \def\Ppi         {\ensuremath{\pi}\xspace}                 
                  
 \def\Prho        {\ensuremath{\rho}\xspace}

 \mathchardef\PDelta="7101
 \mathchardef\PXi="7104
 \mathchardef\PLambda="7103
 \mathchardef\PSigma="7106
 \mathchardef\POmega="710A
 \mathchardef\PUpsilon="7107
                  
 \def\PB      {\ensuremath{B}\xspace}                 
                  
 \def\PD      {\ensuremath{D}\xspace}

 \def\PK      {\ensuremath{K}\xspace}

 \def\Pi      {\ensuremath{i}\xspace}

}

\makeatletter
\ifcase \@ptsize \relax
  \newcommand{\miniscule}{\@setfontsize\miniscule{4}{5}}
\or
  \newcommand{\miniscule}{\@setfontsize\miniscule{5}{6}}
\or
  \newcommand{\miniscule}{\@setfontsize\miniscule{5}{6}}
\fi
\makeatother

\DeclareRobustCommand{\optbar}[1]{\shortstack{{\miniscule (\rule[.5ex]{1.25em}{.18mm})}
  \\ [-.7ex] $#1$}}

\def\pion   {{\ensuremath{\Ppi}}\xspace}

\def\pip    {{\ensuremath{\pion^+}}\xspace}
\def\pim    {{\ensuremath{\pion^-}}\xspace}

\def\rhomeson {{\ensuremath{\Prho}}\xspace}
\def\rhoz     {{\ensuremath{\rhomeson^0}}\xspace}

\def\kaon    {{\ensuremath{\PK}}\xspace}
  \def\Kbar    {{\kern 0.2em\overline{\kern -0.2em \PK}{}}\xspace}

\def\KorKbar    {\kern 0.18em\optbar{\kern -0.18em K}{}\xspace}

\def\Kp      {{\ensuremath{\kaon^+}}\xspace}
\def\Km      {{\ensuremath{\kaon^-}}\xspace}

  \def\Dbar    {{\kern 0.2em\overline{\kern -0.2em \PD}{}}\xspace}
\def\D       {{\ensuremath{\PD}}\xspace}

\def\DorDbar    {\kern 0.18em\optbar{\kern -0.18em D}{}\xspace}
\def\Dz      {{\ensuremath{\D^0}}\xspace}
\def\Dzb     {{\ensuremath{\Dbar{}^0}}\xspace}

\def\Bbar    {{\ensuremath{\kern 0.18em\overline{\kern -0.18em \PB}{}}}\xspace}

\def\BorBbar    {\kern 0.18em\optbar{\kern -0.18em B}{}\xspace}

  \def\Y#1S{\ensuremath{\PUpsilon{(#1S)}}\xspace}

\def\Lbar        {{\ensuremath{\kern 0.1em\overline{\kern -0.1em\PLambda}}}\xspace}
\def\LorLbar    {\kern 0.18em\optbar{\kern -0.18em \PLambda}{}\xspace}

\newcommand{\decay}[2]{\ensuremath{#1\!\to #2}\xspace}         
\def\ra                 {\ensuremath{\rightarrow}\xspace}
\def\to                 {\ensuremath{\rightarrow}\xspace}

\def\order   {{\ensuremath{\mathcal{O}}}\xspace}

\def\CP                {{\ensuremath{C\!P}}\xspace}

\def\AT#1     {\ensuremath{A_{\mathrm{T}}^{#1}}\xspace}           

\def\C#1      {\ensuremath{\mathcal{C}_{#1}}\xspace}                       
\def\Cp#1     {\ensuremath{\mathcal{C}_{#1}^{'}}\xspace}                    
\def\Ceff#1   {\ensuremath{\mathcal{C}_{#1}^{\mathrm{(eff)}}}\xspace}        
\def\Cpeff#1  {\ensuremath{\mathcal{C}_{#1}^{'\mathrm{(eff)}}}\xspace}       
\def\Ope#1    {\ensuremath{\mathcal{O}_{#1}}\xspace}                       
\def\Opep#1   {\ensuremath{\mathcal{O}_{#1}^{'}}\xspace}                    

\newcommand{\tev}{\ifthenelse{\boolean{inbibliography}}{\ensuremath{~T\kern -0.05em eV}\xspace}{\ensuremath{\mathrm{\,Te\kern -0.1em V}}}\xspace}
\newcommand{\gev}{\ensuremath{\mathrm{\,Ge\kern -0.1em V}}\xspace}
\newcommand{\mev}{\ensuremath{\mathrm{\,Me\kern -0.1em V}}\xspace}
\newcommand{\kev}{\ensuremath{\mathrm{\,ke\kern -0.1em V}}\xspace}
\newcommand{\ev}{\ensuremath{\mathrm{\,e\kern -0.1em V}}\xspace}
\newcommand{\gevc}{\ensuremath{{\mathrm{\,Ge\kern -0.1em V\!/}c}}\xspace}
\newcommand{\mevc}{\ensuremath{{\mathrm{\,Me\kern -0.1em V\!/}c}}\xspace}
\newcommand{\gevcc}{\ensuremath{{\mathrm{\,Ge\kern -0.1em V\!/}c^2}}\xspace}
\newcommand{\gevgevcccc}{\ensuremath{{\mathrm{\,Ge\kern -0.1em V^2\!/}c^4}}\xspace}
\newcommand{\mevcc}{\ensuremath{{\mathrm{\,Me\kern -0.1em V\!/}c^2}}\xspace}

\def\order{{\ensuremath{\cal O}}\xspace}
\newcommand{\chisq}{\ensuremath{\chi^2}\xspace}

\def\gsim{{~\raise.15em\hbox{$>$}\kern-.85em
          \lower.35em\hbox{$\sim$}~}\xspace}
\def\lsim{{~\raise.15em\hbox{$<$}\kern-.85em
          \lower.35em\hbox{$\sim$}~}\xspace}

\def\tell1  {TELL1\xspace}
\def\ukl1   {UKL1\xspace}

\newcommand{\eg}{\mbox{\itshape e.g.}\xspace}
\newcommand{\ie}{\mbox{\itshape i.e.}\xspace}

\usepackage{cite} 
\usepackage{mciteplus}

\usepackage{longtable} 

\begin{document}

\renewcommand{\thefootnote}{\fnsymbol{footnote}}
\setcounter{footnote}{1}


%
%


\begin{titlepage}
\pagenumbering{roman}

\vspace*{-1.5cm}
\vspace*{1.5cm}
\noindent
\begin{tabular*}{\linewidth}{lc@{\extracolsep{\fill}}r@{\extracolsep{0pt}}}
\\
 & & \today \\ 
 & & \\
\end{tabular*}

\vspace*{4.0cm}

{\bf\boldmath\huge
\begin{center}
  On model-independent searches for direct \CP violation in multi-body decays
\end{center}
}

\vspace*{2.0cm}

\begin{center}
Chris Parkes, Shanzhen Chen, Jolanta Brodzicka,\\Marco Gersabeck, Giulio Dujany, William Barter

\vspace{0.5cm}

School of Physics and Astronomy, University of Manchester, \\
Oxford Road,
Manchester, M13 9PL, UK
\end{center}

\vspace{\fill}

\begin{abstract}
  \noindent
  Techniques for performing model-independent searches for direct \CP
  violation in three and four-body decays are discussed.  Comments on 
  the performance and the optimisation of a binned 
 \chisq approach and an unbinned approach, known as the energy test,
 are made. The use of the energy test in the presence of background is also studied. The selection and
  treatment of the coordinates used to describe the phase-space of the
  decay are discussed. The conventional model-independent techniques, which
  test for $P$-even \CP violation, are modified to create a new approach
  for testing for $P$-odd \CP violation.  An implementation of the
 energy test using GPUs is described.
  
\end{abstract}

\vspace*{2.0cm}

\begin{center}
\end{center}

\vspace{\fill}

\vspace*{2mm}

\end{titlepage}


\newpage
\setcounter{page}{2}
\mbox{~}
%
%
%
%

\cleardoublepage


\renewcommand{\thefootnote}{\arabic{footnote}}
\setcounter{footnote}{0}



\pagestyle{plain} 
\setcounter{page}{1}
\pagenumbering{arabic}


%


\section{Introduction}
\label{sec:Introduction}

The study of Charge-Parity (\CP) violation allows for a sensitive
probe of new physics from beyond the Standard Model of particle
physics. \CP violation is incorporated in the Standard Model of particle
physics through a complex phase in the
Cabibbo-Kobayashi-Maskawa (CKM) matrix~\cite{Kobayashi:1973fv}. Contributions from new
particles, at mass scales that cannot be directly probed, can enhance
the amount of \CP violation observed. 

This paper discusses model-independent searches for direct \CP violation, \ie\ \CP violation in decays, in
multi-body final states. The rich phase-space of interfering resonances in
multi-body decays provides excellent opportunities for \CP violation
measurements.  Two techniques that have been used in the literature
are discussed in Sect.~\ref{sec:Techniques}.  These are a
binned \chisq test and an unbinned technique based on the average 
weighted distance between events in phase-space known as the
energy test. The performance and optimisation of these techniques 
is discussed. An approach to study \CP violation in the presence of
background events using unbinned techniques is presented in Sect.~\ref{sec:Bkg} through the extension of the energy test formalism.

The selection of the coordinates used for analysing the \CP violation are discussed 
in Sect.~\ref{sec:Coordinate}. A novel method for analysing $P$-odd \CP violation, which is accessible in any decay for which a parity violating observable can be defined such as decays to four pseudo-scalar particles, is presented in Sect.~\ref{sec:POdd}.
This method is applicable for use with
any two sample comparison test. The first application of this
technique~\cite{OurNewPaper}, using the energy test to compare the samples,  
has given rise to a result 2.7$\sigma$ away from the no-\CP violation hypothesis.

The implementation of the energy test method in a
computationally efficient manner using GPUs is discussed in
Sect.~\ref{sec:GPU}.

\section{Techniques}
\label{sec:Techniques}
Model-independent searches for direct \CP violation have been carried out
using a number of different techniques. Tests have often been 
performed using a binned \chisq  approach to compare the relative density in the Dalitz plot~\cite{Dalitz:1953cp} of
a decay and its \CP-conjugate sample (see for example
\cite{Aubert:2008yd} for three-body and \cite{Aaij:2013jxa} for
a generalisation to the phase-space of four-body decays). This method is discussed further
below. More recently unbinned techniques have also been applied. A technique known as the energy test
has been applied (see~\cite{LHCb-PAPER-2014-054} for three-body and
~\cite{OurNewPaper} for four-body), again this is discussed
below. A nearest neighbours approach has been used in Ref.~\cite{LHCb-PAPER-2013-057}, albeit with a very small number of neighbours, 
and the angular moments of the cosine of the helicity angle of the
studied particle have also been utilised~\cite{Aubert:2008yd} in three-body decays.
A class of measurements based on triple products of momenta in
four-body meson decays have also been performed \cite{Link:2005th}.
The $P$-odd \CP violation test proposed in Sect.~\ref{sec:POdd} is related to these.
$P$-odd \CP violation can also be accessed in the baryon sector owing to significant $P$-parity violation 
even in two-body decays of baryons. This type of \CP asymmetry can be measured by comparing $P$ asymmetries 
for baryon and anti-baryon decays~\cite{Link:2005ft}.

\subsection{Binned \chisq}
\label{sec:Scp}

The simplest and most commonly applied technique used in the literature, here referred to
as $S_{\CP}$~\cite{Aubert:2008yd,Bediaga:2009tr},  is a two-sample
binned $\chisq$ test. The phase-space is divided into bins. The statistical significance
of the difference in number of entries in the bin for the $X$ and
$\bar{X}$ samples is computed.
\begin{equation}
S^i_{\CP}= \frac{N^i (X) - \alpha N^i (\bar X)}{\sqrt{N^i (X) + \alpha^2 N^i (\bar X)} }, \quad
\alpha = \frac{N_{tot}(X)}{N_{tot}(\bar X)}, 
\end{equation}
where $N^i (X)$ and $N^i (\bar X)$ are the numbers of $X$ and $\bar X$
candidates in the $i^{\rm th}$ bin, and the $N^i$ values are sufficiently
large that Gaussian uncertainties may be assumed. $\alpha$ is the ratio between the total
yield of $X$ and $\bar X$ events. The parameter $\alpha$ is introduced
to account for global asymmetries which may occur due to production effects. The small correction to the significance
term in the denominator varies in the literature \cite{Aubert:2008yd,
  Aaij:2013jxa} where the form given here is recommended.

In the absence of local \CP asymmetries the $S^i_{\CP}$ are distributed
according to a Gaussian of unit width and zero mean. The $\chisq$ test
value is computed from $\chisq = \sum{(S^i_{\CP})^2}$.
The corresponding $p$-value for the compatibility of the observed data
with the no \CP violation hypothesis can be computed directly from the
observed $\chisq$ value and the number of degrees of freedom; here equal to number of bins$-1$. The test
is  straightforward to implement and requires only minimal
computing resources.

The number, size and location of the bins need to be selected by the
analysts. General advice for $\chisq$ comparison tests is that the number of bins must be sufficient not to miss local 
regions of asymmetry but limited to ensure sufficient numbers of entries in the bins to not
affect the sensitivity. 

The number of bins used in the method in the literature have varied significantly.
The initial application and discussion of the method  \cite{Aubert:2008yd,Bediaga:2009tr}
divided the Dalitz plane into $\order(10^3)$ bins. Applications of
the method have also used a smaller number of bins, $\order(10)$ to
$\order{(10^2)}$ \cite{Aaij:2011cw,Aaij:2013jxa}.  

\begin{figure*}[bt]
    \centering
       \includegraphics[width=0.49\textwidth]{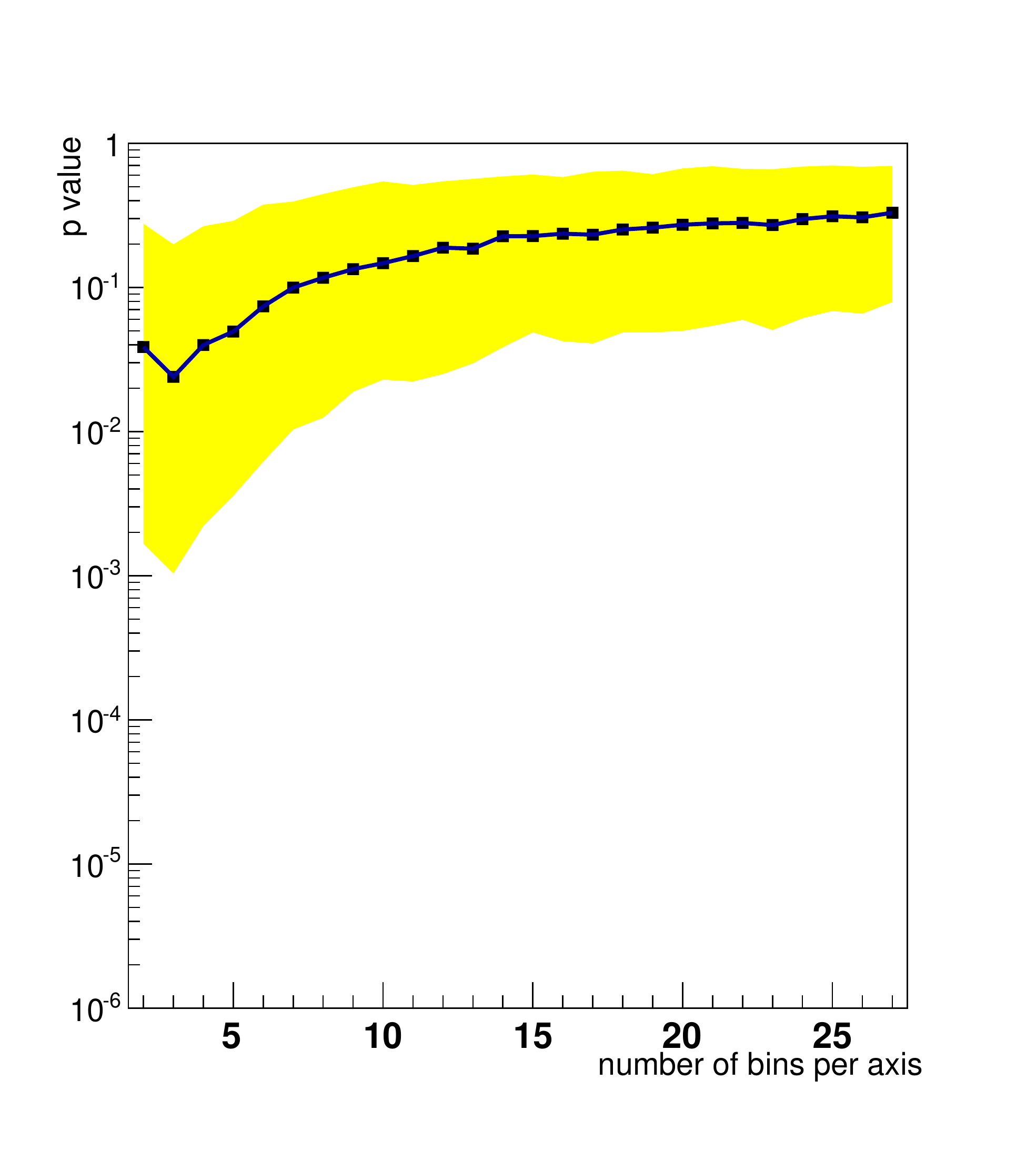}
       \includegraphics[width=0.49\textwidth]{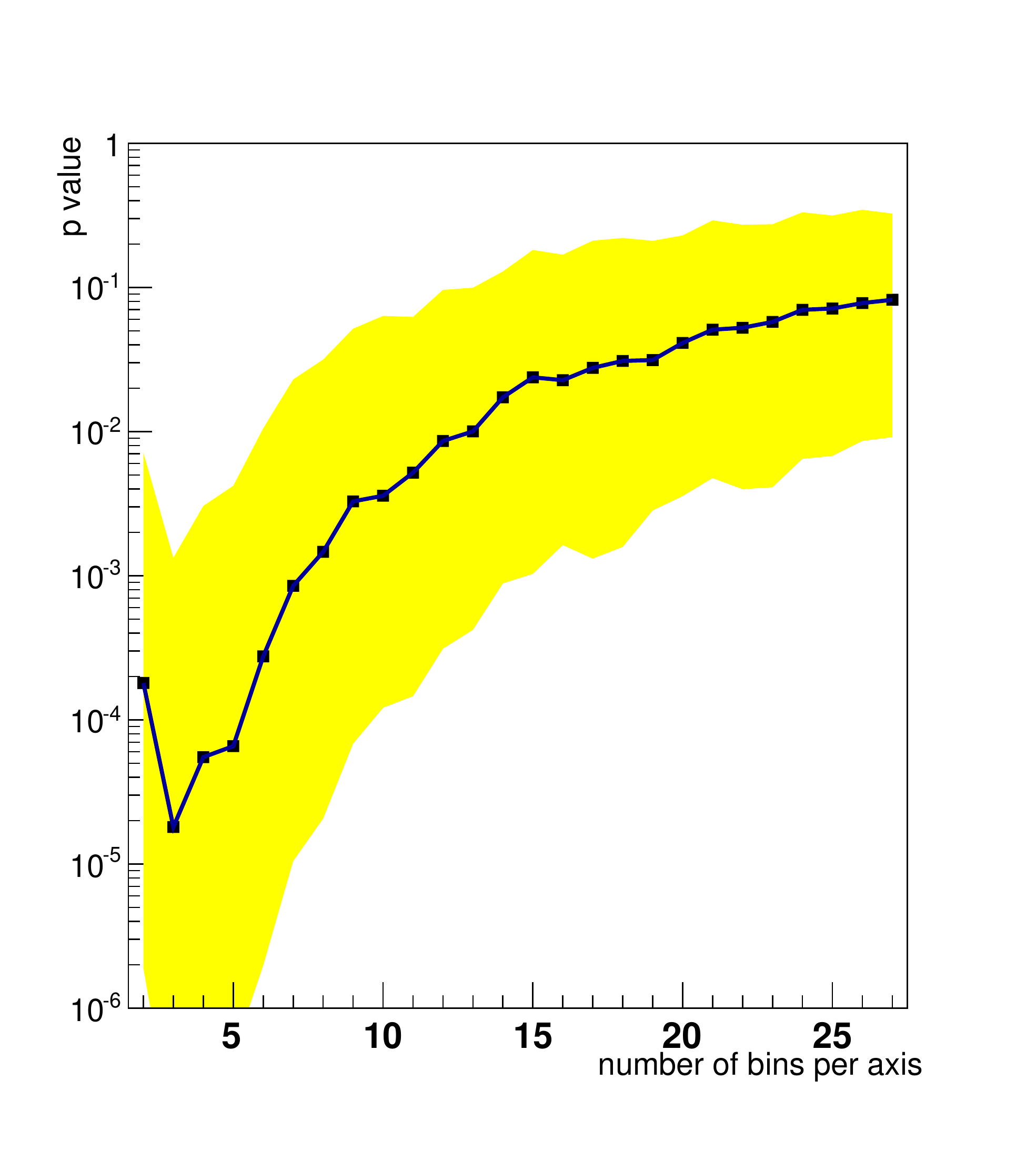}      
\\
     \caption{
       \small $p$-value versus number of bins in $S_{\CP}$ method for 
       simulation samples with \CP violation introduced at the (left)
       3\% and (right) 5\%
       level in one amplitude (see text). 1000 samples were generated and the mean $p$-values are
       shown by the points and the one-sigma confidence level range
       indicated by the yellow-band.
       }
     \label{fig:SCPBins}
    \end{figure*}

We recommend that the number of bins used in the method should be kept
to a relatively small number. The number of degrees of freedom
increases for every additional bin used and consequently the
sensitivity of the method is decreased. This is illustrated in
Fig.~\ref{fig:SCPBins}, where the increased sensitivity of using a
smaller number of bins is clearly observed.  In this study simulation samples  
of 100,000 events were generated using the analysis package Laura++~\cite{Laura} according to the
simple model in Ref.~\cite{Williams:2011cd}. \CP violation was introduced by
changing the amplitude of the resonance with the largest fraction in that model. 

In a binned approach there is
a clear trade-off between minimising the number of bins used and
retaining sensitivity to the rich phase-space of interfering
amplitudes in the decay. This is particularly true in the case of
four-body decays where five coordinates are required to describe the
phase-space (see Sect.~\ref{sec:Coordinate}).
 
Binned distributions of the \CP asymmetry in the phase-space can also be
used to test for \CP violation. This technique was successfully applied
in \cite{Aaij:2013sfa, Aaij:2013bla} to observe local \CP asymmetries in
$B^+\rightarrow h^+h^+h^-$ decays with $h=\pi, \ K$. In this application of $\order(10^2)$
bins were used.
The placement of the bins was physically motivated by the observed
location of the resonances. The bin sizes varied across the plane to
equalise the number of entries in each bin.

\subsection{Energy test}
\label{sec:EnergyTest}

A wide range of unbinned two sample tests exist in the
literature~\cite{Porter:2008mc}. Well-known examples include
the nearest neighbour approach, which has previously been applied to
\CP violation tests~\cite{LHCb-PAPER-2013-057}, and multi-dimensional
extensions of the Kolmogorov-Smirnov test, commonly used in the
astronomy literature~\cite{Fasano:1987}. The former has previously been
shown to be relatively
insensitive for this class of problem~\cite{Williams:2010vh} and studies
by the authors have concluded similarly for the latter.

The class of multi-variate tests based on distances between observables are of
particular interest.
A statistical method called the energy test was introduced in Refs.~\cite{doi:10.1080/00949650410001661440,Aslan2005626}.
Reference~\cite{Williams:2011cd} suggests applying this method to Dalitz plot analyses
and demonstrates the potential to obtain improved sensitivity to \CP violation over
the standard binned approach. The distribution of the test statistic
is not known, and the permutation method is required to be applied in
order to determine the significance of a result. This method was applied
in~\cite{LHCb-PAPER-2014-054} and is described below and used in the studies
in this paper. Many other tests of this class are known, including
cases where the exact
distribution is known under the null-hypothesis, and hence no
permutations would be required (\eg Cross-match statistic test~\cite{Rosenbaum}).
 
In the energy test method a test statistic, $T$, is used to compare average distances in phase-space,
based on a distance function, $\psi_{ij}$, of pairs of events $ij$
belonging to two samples. In the standard test the two samples are
those of different flavours, particle and anti-particle. 
The test statistic is defined as
\begin{equation}
T = \sum_{i,j>i}^{n}\dfrac{\psi_{ij}}{n(n-1)}
+ \sum_{i,j>i}^{\overline{n}}\dfrac{\psi_{ij}}{\overline{n}(\overline{n}-1)}
- \sum_{i,j}^{n,\overline{n}}\dfrac{\psi_{ij}}{n\overline{n}} ,
\label{eqn:T}
\end{equation}
where the first and second terms correspond to a weighted average distance
of events within the $n$ events of the first sample and the
$\overline{n}$ events of the second sample, respectively.
The third term measures the weighted average distance of events in one flavour sample to events
of the opposite flavour sample. The normalisation factors in the denominator
remove the impact of global asymmetries between the two samples.

If the distributions of events in both samples are identical the measured $T$ value 
will fluctuate around a value close to zero; differences between these distributions 
increase the value of $T$. This is translated into a $p$-value under the hypothesis of \CP symmetry 
by comparing the measured $T$ value to a distribution of $T$ values obtained from permutation samples. 
The permutation samples are constructed by randomly assigning events to either of the samples, 
thus simulating a situation without \CP violation. The $p$-value for the no-\CP-violation hypothesis 
is obtained as the fraction of permutation $T$ values greater than the observed $T$ value.

If large \CP violation is observed, the observed $T$ value is likely to lie outside the range of permutation 
$T$ values. In this case the permutation $T$ distribution can be fitted with a generalised-extreme-value (GEV) 
function, as demonstrated in Refs.~\cite{doi:10.1080/00949650410001661440,Aslan2005626} and used 
in Ref.~\cite{LHCb-PAPER-2014-054,OurNewPaper}. The $p$-value from the fitted $T$ distribution can 
be calculated as the fraction of the integral of the function above the observed $T$ value.

The distance function should be falling with increasing distance $d_{ij}$ between events $i$ and $j$,
in order to increase the sensitivity to local asymmetries. A Gaussian function is chosen, defined as
$\psi_{ij}\equiv\psi(d_{ij})=e^{-d_{ij}^2/2\sigma^2}$ with a tunable parameter $\sigma$, which
describes the effective radius in phase-space within which
a local asymmetry is measured. Thus, this parameter should be larger than the resolution of $d_{ij}$
but small enough not to dilute locally varying asymmetries.

\begin{figure*}[bt]
    \centering
       \includegraphics[width=0.7\textwidth]{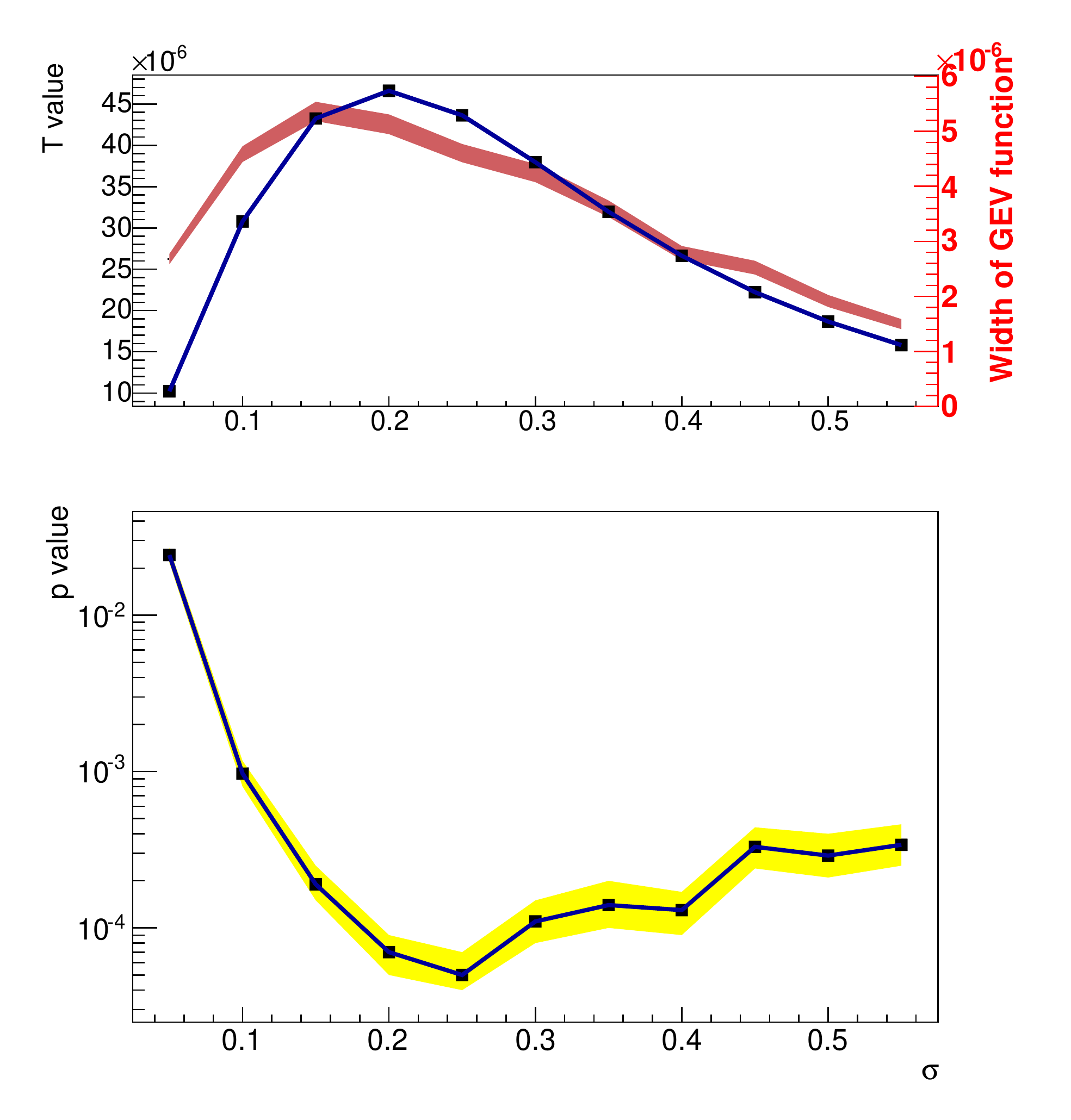}\\
     \caption{
       \small (top) T-values and (bottom) $p$-values versus $\sigma$ in the energy test method
       for simulation samples with \CP violation introduced at the 3\%
       level in one amplitude (see text). The $p$-values are evaluated
       using 100 permutation samples where their distribution is
       fitted with a GEV function. The width of the GEV function and
       its uncertainty is also shown on the top plot. The yellow band
       on the lower plot shows the uncertainty on the $p$-value
       accounting for the uncertainties in the GEV fit, including the
       correlation of the fitted parameters.}
     \label{fig:EnergySigma}
    \end{figure*}

The performance of the energy test method, based on one of the 
samples used in Fig.~\ref{fig:SCPBins}, is shown in
Fig.~\ref{fig:EnergySigma}. The sensitivity to the parameter $\sigma$
is shown. The improved performance of this method over $S_{\CP}$ in
this case of a 3\% asymmetry in amplitude is seen by comparing the
figures; indeed, a 5\% amplitude asymmetry results in $p$-values from the energy test
below $10^{-10}$. Studies from the
authors show this enhancement in sensitivity is common but not
universal: cases have been found in three-body decays
in which $S_{\CP}$ showed better sensitivity than the energy test for
optimal binning and \CP violation in the amplitude or phase of some
resonances.
The uncertainty on the $p$-value in Fig.~\ref{fig:EnergySigma} is obtained by randomly 
resampling the fit parameters within their uncertainties, taking into account their correlations, 
and by extracting a $p$-value for each of these generated $T$ distributions. This uncertainty
is found to arise mainly from the uncertainty on the width of the GEV function 
(Fig.~\ref{fig:EnergySigma}). 

The sensitivity of the optimal $\sigma$ to various parameters of a
three-body decay model was studied. The value was found to be largely independent of the Dalitz
plot structure or the width of the resonance in which \CP violation
was introduced. A variable $\sigma$ set according to the density of
points in the plot was also studied but not demonstrated to have any
advantage over a fixed $\sigma$. As trivially expected, the optimal $\sigma$ is
dependent quadratically on the mother particle mass in a three-body decay.

 \section{Presence of background}
 \label{sec:Bkg}
 \CP violation is expected to appear in relatively low rate decays where two amplitudes of similar magnitude interfere. The samples of signal events under study may be significantly polluted by additional background events from different physics processes that contribute at a similar rate to signal. Such pollution reduces the fraction of signal events (the purity) of the sample under study, and requires the use of additional techniques such that any significant effect can be associated with \CP violation in the signal process.

 The presence of background is problematic when studying local \CP
 violation in the case that the purity is different in the two samples that are
 being compared. This can be generated, for example, through the presence of global production asymmetries between the two samples that are different for signal and background, even if no \CP violation is present. Similarly problematic is the case where the background process itself exhibits \CP violation. In these scenarios the local densities of events will differ between the two samples, since background populates the phase space differently to signal. For the binned $S_{\CP}$ test different yields will be found in the phase space bins (even after normalising for the overall yield), and a significant effect could be inferred. The effect of background on the $S_{\CP}$ test is typically removed by fitting the invariant mass distribution in each bin to determine the number of signal events in each bin, and then comparing these between the two samples. The energy test also exhibits similar complications. Differences in purity between the two samples are not reproduced when events are randomly assigned a flavour, since the purities and event densities will be consistent for the permuted samples; the densities of samples used in the permutation studies will not accurately model differences in density between the two `real' data samples. Consequently background events can lead to significant $T$ values and $p$-values being calculated even when the signal process exhibits no \CP violation.  It is therefore important to consider techniques to remove the effect of background from the energy test, so that any significant $T$ value can be associated with \CP violation in the studied decay. This section sets out such an approach.
 
The weighting of events to remove the effect of background was
considered in Ref.~\cite{Williams:2011cd}. Such weights give the
relative purity of each sample in the particular region of phase-space
that the event is drawn from. Therefore, when $T$ is calculated two
weights are used for each pair of events, with one weight for each event. This allows each pair of events to be weighted according
to the relative number of pairs of signal events expected to
contribute to the calculation of $T$ from such locations within the
phase space, and allows the estimation of the $T$ value that would be
found if only signal were present. However, this approach relies on
knowledge of the signal density within the phase-space, which in many practical
cases is not known.

An alternative method is suggested here to remove the bias introduced by
background if additional representative samples of background events are available,
for example from signal side-bands. This method will be equally
applicable to other unbinned model independent two sample tests but is
discussed here for the energy test. These additional samples will be labelled here as `background samples', as opposed to the `main samples' which are being explicitly tested for differences in local event densities. The events in the background samples can be used to subtract off the effect of signal and background event pairs, and background and background event pairs that arise when considering the $T$ value set out in Eq.~\ref{eqn:T} when background is present in the main samples. This can be achieved by altering the test statistic to
\begin{align}
  T &= \frac{1}{2w(w-1)}(\sum\limits_{i}^{n}\sum\limits_{j \neq i}^{n} \psi_{ij}  - \frac{2b}{b_s}\sum\limits_{i}^{n}\sum\limits_{j}^{b_s} \psi_{ij}  + \frac{b(b+1)}{b_s(b_s-1)}\sum\limits_{i}^{b_s}\sum\limits_{j\neq i}^{b_s} \psi_{ij} )\nonumber\\
  &+\frac{1}{2\bar{w}(\bar{w}-1)}(\sum\limits_{i}^{\bar{n}}\sum\limits_{j \neq i}^{\bar{n}} \psi_{ij}  - \frac{2\bar{b}}{\bar{b_s}}\sum\limits_{i}^{\bar{n}}\sum\limits_{j}^{\bar{b_s}} \psi_{ij}  + \frac{\bar{b}(\bar{b}+1)}{\bar{b_s}(\bar{b_s}-1)}\sum\limits_{i}^{\bar{b_s}}\sum\limits_{j\neq i}^{\bar{b_s}} \psi_{ij} )\nonumber\\
  &-\frac{1}{w \bar{w}}(\sum\limits_{i}^{n}\sum\limits_{j}^{\bar{n}} \psi_{ij}  - \frac{\bar{b}}{\bar{b_s}}\sum\limits_{i}^{n}\sum\limits_{j}^{\bar{b_s}} \psi_{ij}  - \frac{b}{b_s}\sum\limits_{i}^{b_s}\sum\limits_{j}^{\bar{n}} \psi_{ij}  + \frac{b\bar{b}}{b_s\bar{b_s}}\sum\limits_{i}^{b_s}\sum\limits_{j}^{\bar{b_s}} \psi_{ij} ),
  \label{eqn:t_bkg}
  \end{align} 
where $w$ and $\bar{w}$ are the number of signal events in the main
samples, and $b$ and $\bar{b}$ are the number of background events in
the main samples, while $b_s$ and $\bar{b_s}$ denote the number of
background events in the background samples. The additional terms in
comparison with Eq.~\ref{eqn:T} sum over pairs of events in the
background and main samples or sum over pairs of events in the
background samples, removing on average the effects of pairs of
background and signal events and pairs of background events, when
calculating $T$. The other notable difference with Eq.~\ref{eqn:T} is
the inclusion here also of terms with $j<i$ when the sums are over the same sample. This is balanced with an additional factor of $\frac{1}{2}$, and is included for simplicity when considering all terms.
Assuming that the density of events in the background samples reflects the density of background events in the main samples (up to an overall normalisation factor), this $T$ value provides an unbiased estimate of the $T$-value that would be calculated in the presence of signal alone.

\begin{figure*}[bt]
    \centering
       \includegraphics[width=0.6\textwidth]{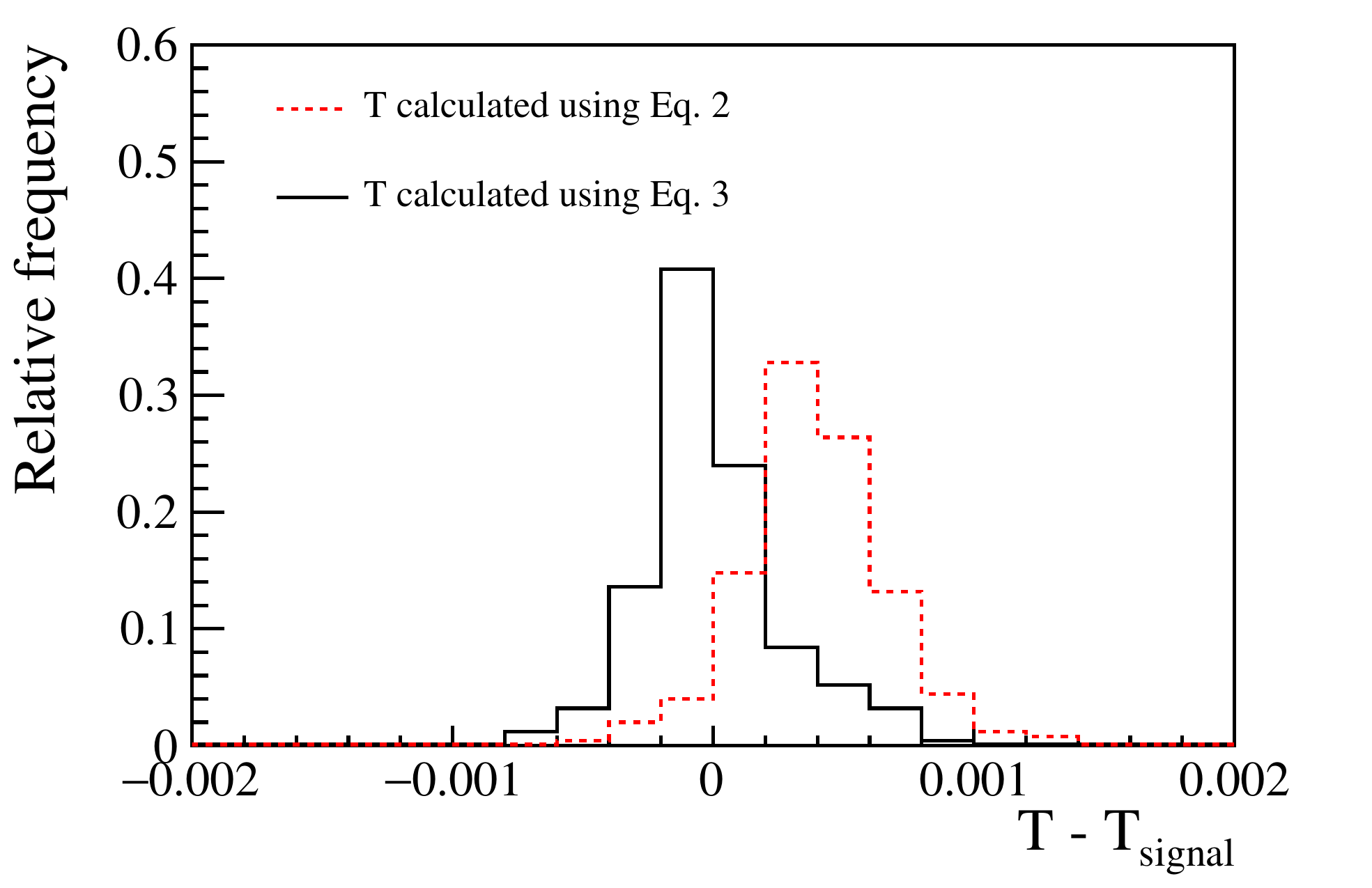}
\\
     \caption{
       \small The $T$-value relative to the $T$-value found from only considering signal events ($T_\text{signal}$), when calculated using Eq.~\ref{eqn:T}, where background is not removed, and Eq.~\ref{eqn:t_bkg}, which subtracts (on average) the effect of background. The first case shows a clear bias; the presence of background can lead to small $p$-values when the permutation method is used to determine significance. The mean of the second distribution is consistent with zero, removing the bias associated with background. 
       }
     \label{fig:Background}
    \end{figure*}

An example of the effect of background on the energy test is shown in
Fig.~\ref{fig:Background}. Here, 4,000 signal events were generated
(without $CP$ violation) using the same model as in
Ref.~\cite{Williams:2011cd}, and assigned a flavour randomly, with a 50\%
chance of each flavour. Background events were generated assuming no
variation in event density across the 3-body phase space, with 1,600
events contaminating the main samples. Again, the flavour was assigned
randomly, though on average 75\% of events were assigned one flavour
and 25\% the other. An additional 1,600 events (with the same 3:1
asymmetry in flavour) were used to create additional background
samples that could be used to remove any bias introduced by background
in the main samples. This was performed 250 times, and in each trial
the energy test was calculated using a $\sigma$ parameter of
0.25. Such a scenario can generate large $T$-values if Eq.~\ref{eqn:T}
is used to calculate the $T$-value, and the permutation method yields
significant $p$-values. However, if the background samples are also
considered, and Eq.~\ref{eqn:t_bkg} is used to remove the effects of
background on average, an unbiased estimate of the $T$-value of signal
events is recovered. This allows the use of the permutation method to
estimate $p$ values (using the same method to remove background from
the randomly permuted samples), and allows any significant effect to
be associated clearly with the signal channel. Therefore, for the rest
of this article, the case where background is present is neglected.

\section{Coordinate selection}
\label{sec:Coordinate}

A decay of a pseudo-scalar particle $M$ into $n$ pseudo-scalar particles ($A,B,C...$) $M\ra ABC...(n)$ 
can be described by $n$ four-vectors $p^\mu_A, p^\mu_B, p^\mu_C...$,
and consequently $4n$ parameters. The known masses of the identified
final state particles $A,B,C...$ remove $n$ degrees of freedom. E,$\bf p$
conservation removes an additional four degrees of freedom. The system can be
rotated freely around all spatial axes, removing a further three
degrees of freedom. Hence, $3n-7$ degrees of freedom remain. Consequently a
three-body decay phase-space is fully described by the two variables conventionally
used in Dalitz plot analyses. The phase-space of a four-body decay can be fully described by 
five parameters. The selection of the variables used to describe the phase-space is
discussed in this section.

\subsection{Distances in Phase-Space}
\label{sec:Distances}

The $S_{\CP}$ method requires the division of the decay phase-space
into bins. The energy test method relies on the distance between
events in the phase-space. The result of model-independent
two-sample comparison tests will typically depend on the distance
metric, not only the coordinates chosen.

The distance between two points in phase-space in a three-body decay is usually measured as the 
Euclidean distance in the Dalitz plot.
However, this distance depends on the choice of the axes of the Dalitz plot.
This dependence can be removed by using all three invariant masses to determine
the distance, $d_{ij}$, calculated as the length of the displacement vector
$\Delta\vec{x}_{ij}=(m_{12}^{2,j}-m_{12}^{2,i},m_{23}^{2,j}-m_{23}^{2,i},m_{13}^{2,j}-m_{13}^{2,i})$,
where the $1,2,3$ subscripts indicate the final-state particles.
Using all three invariant masses does not add information,
but it avoids an arbitrary choice that could impact the sensitivity of the method to different
\CP violation scenarios. This symmetrisation was applied in Ref.~\cite{LHCb-PAPER-2014-052}.

In describing a four-body decay phase-space with invariant masses a mixture of two
and three-body masses may be preferred, with no
directly analogous symmetrisation, as discussed below. 

\subsection{Coordinates in four-body decays}
\label{sec:CoordinatesFourBody}

Four-body decays offer both challenges and opportunities with respect to three-body decays.
A four-body decay phase-space is obviously more complicated and five coordinates are needed 
for its full description~\cite{Byckling}. No unique choice of variables exists and depending 
on the decay dynamics and the purpose of the measurement one can try and
optimize the set of coordinates. 

For cascade-type decays proceeding via resonances in a three-body subsystem and 
followed by two-body decays (e.g. $D^0 \to a_1(1260)^+ (\to \rhoz(770) \pi^+) \pi^-$) 
a three-body invariant mass and a Dalitz-style distribution of the three-body decay would
be a preferred option.  
For decays occurring through two two-body resonances, in particular ones with non-zero spins, 
(e.g. $D^0\to\rhoz(770) \rhoz(770)$) the natural choice is the so called transversity basis~\cite{focus} 
comprising invariant masses of two-body subsystems, their helicity angles 
and the acoplanarity angle between decay planes of the two resonances. 
In some decays both decay types may contribute significantly (\eg\ both
of the examples given here contribute to \decay{\Dz}{\pip\pim\pip\pim} decays).
The most general amplitude analysis that aims at a description of decay dynamics is performed 
in the phase-space constructed with the four-momenta of the final state particles and 
any coordinates are chosen only to illustrate the results.

The energy test is a statistical method comparing the 
events distributions in phase-space (see Sect.~\ref{sec:Techniques}). 
Therefore it is sensitive to the position of an event 
in phase-space and to the choice of coordinates building this phase-space.
All choices are not equivalent in terms of the sensitivity of the analysis 
as it will change the distance between events in the phase-space. 

There are many possible choices for the variables to describe the
degrees of freedom of the phase-space. A natural set of invariants is $p_ip_j$, or equivalently $s_{ij}$
defined as $s_{ij} = (p_i + p_j)^2 = m_i^2 +m_j^2 +2(p_i p_j)$.
There are $\frac{1}{2} n(n-1)$ such invariants; for four-body decays there are six of them.

Three-body mass invariants, $s_{ijk}$, can be made from
linear combinations of the two-body invariants: 
$s_{ijk} = (p_i +p_j +p_k)^2 = s_{ij} +s_{ik}+s_{ik} - m_i^2 -m_j^2-m_k^2$,
so can be used interchangeably (for $n>3$) for the $s_{ij}$.
While they carry the same information as $s_{ij}$, the choice will change the distance
measure between events used in the \CP violation search.
There are $\frac{1}{3!} n(n-1)(n-2)$ such invariants; there are four of them in four body
decays.

Consequently there are ten mass invariants, six two-body and four three-body, which
can be used to characterise a four-body decay. We consider a
physically motivated choice of the coordinates. Invariant mass
combinations corresponding to `unphysical' doubly-charged meson
resonances may be excluded. A further reduction in coordinates can be
made by excluding mass combinations with small resonance
contributions. 
 
An additional complication in many decays of interest will be the presence of identical
final state particles (e.g. \decay{\Dz}{\pip\pim\pip\pim}).
Identical particles of the same charge can be swapped; as a result in
the example decay each event  can be placed in four points in phase-space. The energy test is sensitive to such particle swapping 
as well. To get a unique output from the energy test, as well as to
get optimal sensitivity,  the order of the particles, \ie\ the input variables of energy test need to be determined. 

Consider the experimentally interesting example of the singly Cabibbo suppressed \decay{\Dz}{\pip\pim\pip\pim} decay.
The charge order of the particles in the \Dz decay $\pi_1\pi_2\pi_3\pi_4$ is fixed to 
\pip\pim\pip\pim. For a \Dzb decay, the order of $\pi_1\pi_2\pi_3\pi_4$ is the $C$-conjugated one: \pim\pip\pim\pip.
The invariant masses of all possible \pip\pim pairs are calculated and sorted for each event.
Once the \pip\pim pair with the largest invariant mass is fixed to be $\pi_3\pi_4$, 
the order of all four pions is fully determined. 
As only a small fraction of the $\rho(770)$ resonance, either produced directly from \Dz or 
through $a_1(1260)$ decays, contributes to the largest $m(\pip\pim)$, 
the $\pi_3\pi_4$ combination is excluded~\cite{focus}. Two-body masses except for 
the $m(\pi_3\pi_4)$ and three-body mass combinations that do not contain the $\pi_3\pi_4$ are kept. 
In that way we end up with exactly five invariant masses, which contain
most of the dominant resonance contributions, as listed in Table~\ref{tab:PhaseSpaceVariables}. 
This choice has been adopted in~\cite{OurNewPaper}. Simulation
studies comparing the performance of the test with these five
coordinates and with the eight 'physical' mass combinations have been
performed. No significant difference in sensitivity was obtained,
though the optimal sigma is larger when using eight coordinates.

The other singly Cabibbo suppressed \Dz decay with charged
long-lived hadrons in the final state is \decay{\Dz}{\Kp\Km\pip\pim}.
In this case no significant three-body resonances containing \Kp\Km
are observed \cite{Artuso:2012df}, and thus these masses can be excluded as
coordinates. The other coordinates have significant contributions and
thus for an energy test method of this final state a physically
motivated choice would be to proceed with six coordinates, as shown in
Table~\ref{tab:PhaseSpaceVariables}, or to reduce to five coordinates
by further excluding the mass combination with the smallest  resonance
contribution ($m_{234}$).

\begin{table}
 \centering
\begin{tabular}{|c|cc|}
\hline 
 \decay{\Dz}{\pi_1^+\pi_2^-\pi_3^+\pi_4^-}& Two-body masses & Three-body masses \\ 
\hline 
Unphysical & $m_{13},m_{24}$ & \\
\hline 
Physical & $m_{12}, m_{14}, m_{23}, m_{34} $ & $m_{123},m_{124},
                                               m_{134},m_{234}$ \\
\hline
Selected  & $m_{12}, m_{14}, m_{23}$ & $m_{123},m_{124}$ \\
\hline 
\decay{\Dz}{\PK_1^+\PK_2^-\pi_3^+\pi_4^-}& Two-body masses & Three-body masses \\ 
\hline 
Unphysical & $m_{13},m_{24}$ & \\
\hline 
Physical & $m_{12}, m_{14}, m_{23}, m_{34} $ & $m_{123},m_{124},
                                               m_{134},m_{234}$ \\
\hline
Selected  & $m_{12}, m_{14}, m_{23},m_{34}$ & $m_{134},(m_{234})$ \\
\hline 
\end{tabular} 

\caption{Coordinates used, and those excluded, in the measurement of \decay{\Dz}{\pip\pim\pip\pim} 
decays~\cite{OurNewPaper} and coordinates suggested for
\decay{\Dz}{\Kp\Km\pip\pim} decays. Excluding the mass marked in
brackets would reduce to the minimal five coordinates that span the phase-space.}
\label{tab:PhaseSpaceVariables}
\end{table}

\section{Parity-even and Parity-odd \CP Violation Tests}
\label{sec:POdd}

The choice of only invariant masses as coordinates, as discussed in
the previous section, has a limitation. Invariant masses, 
of both two- and three-body systems, can be expressed through the double product 
of particle momenta ($\vec{p}_a \cdot \vec{p}_b$) and, as such, 
are even under the $P$-parity transformation (changing $\vec{p}$ into $-\vec{p}$). 
Thus, invariant masses allow only $P$-even \CP asymmetries to be
probed.  This is also true of helicity angles.

In three-body decays only $P$-even amplitudes can be present and the
conventional model-independent test of comparing particle and
anti-particle samples is sufficient. In four-body decays $P$-odd amplitudes can be present and 
accessed with $P$-odd quantities. These are triple products of a general form 
$\vec{p}_a \cdot (\vec{p}_b \times \vec{p}_c)$; the acoplanarity
angle between decay planes of two resonances is one example. 
There is a class of measurements based on the triple products, 
often called $T$-odd measurements\footnote{Triple products are also
  odd under $T$-parity, although time reversal is not what is typically measured.}, which probe the $P$-odd type of \CP asymmetry only~\cite{LHCb-PAPER-2014-046}.

An asymmetry of this $P$-odd kind is induced by interferences between $P$-even and $P$-odd decay amplitudes, 
thus the sensitivity to \CP violation depends on the $P$-odd amplitude contribution in
the decay. However, it is proportional to the cosine  of the
strong-phase difference between the interfering
partial waves~\cite{Valencia} and thus will be enhanced where $P$-even \CP asymmetry, 
proportional to sine of the strong-phase difference, lacks
sensitivity.

Consider again the general four-body decay \decay{M}{ABCD} and its
antiparticle equivalent $\overline{M}$. A triple product $C_T=\vec{p}_A \cdot (\vec{p}_B \times \vec{p}_C)$ is constructed for $M$ decays. 
The $\vec{p}_A$, $\vec{p}_B$ and $\vec{p}_C$ are vector momenta of particles $A,B,C$ in the $M$ centre-of-mass frame.
The corresponding triple product for the $\overline{M}$ decays is obtained by
applying the \CP parity transformation,
$\CP(C_T)=-C(C_T)=-\overline{C}_T$. The $\overline{C}_T$ is constructed
with the anti-particles $\overline{A},\overline{B},\overline{C}$, being the $C$-conjugations of the ones entering 
$C_T$.
The total sample may be divided into four subsamples according to the
particle/antiparticle flavour and the triple product sign:
\begin{equation}
\label{eqn:tp_split}
[I] \ D(C_T>0), \ \ [II] \ D(C_T<0), \ \ [III] \ \overline{D}(-\overline{C}_T>0), \ \ [IV] \ \overline{D}(-\overline{C}_T<0).
\end{equation}

\begin{figure*}[bt]
    \centering
       \includegraphics[width=0.6\textwidth]{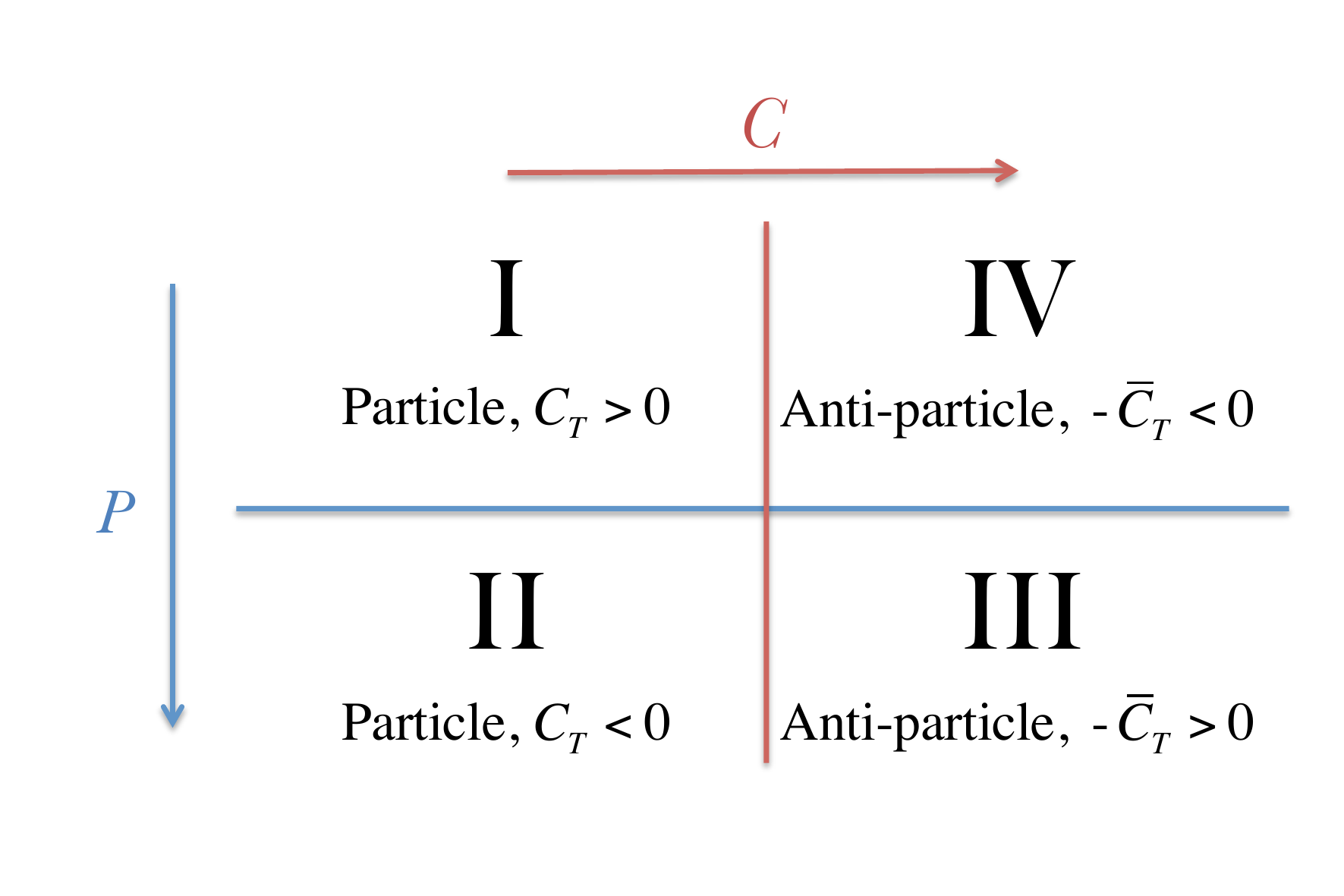}
       \caption{
       \small Symmetry transformation relationships of the four data
       samples used in the \CP violation tests.}
     \label{fig:SamplesSymmetry}
    \end{figure*}

The relationships between the samples under symmetry transformations is
illustrated in Fig.~\ref{fig:SamplesSymmetry}. Samples [I] and [III] are related by \CP transformation; 
and so are [II] and [IV]. There are thus two
potential sample comparison tests for \CP violation using the full
data sample: comparing a sample consisting of data set [I] and [II]
with a sample containing data [III] and [IV]; or comparing the
combined [I]+[IV] with the combined [II]+[III]. Both tests span the full $C_T$ space. 

Consider the more familiar case of  asymmetries, these may be measured in the $C_T$ regions using the number 
of events populating the four samples 
in Eq.~\ref{eqn:tp_split}:
\begin{equation}
\label{eqn:tp_asym}
A_{\CP}(C_T>0) = \frac{N(I)-N(III)}{N(I)+N(III)}, \
A_{\CP}(C_T<0) = \frac{N(II)-N(IV)}{N(II)+N(IV)}.
\end{equation}

In the absence of \CP Violation both of the asymmetries are expected
to be compatible with zero. 

\CP  asymmetries can be extracted from these samples that are $P$-even or $P$-odd (i.e. even and odd with respect to $C_T$)
simply by adding or subtracting the asymmetries measured in the $C_T$ regions:
\begin{equation}
\label{eqn:even_odd_asym}
A_{\CP}^{P-even}=\frac{A_{\CP}(C_T>0) + A_{\CP}(C_T<0)}{2}, \ A_{\CP}^{P-odd}=\frac{A_{\CP}(C_T>0) - A_{\CP}(C_T<0)}{2}.
\end{equation}

The  $A_{\CP}^{P-even}$ test corresponds to integrating  over $C_T$,
and is equivalent (within normalisation) to
the default sample comparison test in which particle (samples [I] and [II]) and
anti-particle samples (samples [III] and [IV]) are compared 
in the phase-space built with invariant masses only. A binned or
unbinned comparison of the phase-spaces of the decays is then
performed using techniques such as those described in Sect.~\ref{sec:Techniques}.

The $A_{\CP}^{P-odd}$ from Eq.~\ref{eqn:even_odd_asym} is equivalent to
the quantity measured in the $T$-odd analyses. The asymmetry is
typically measured integrated over the whole phase-space or
asymmetries can be measured in the phase-space regions~\cite{LHCb-PAPER-2014-046}. 

Unbinned model independent techniques do not allow for an asymmetry measurement. 
However, the $P$-odd \CP asymmetry can be tested by comparing the combined sample $I+IV$ with the combined sample $II+III$. 
This comparison may be performed in the same phase-space as the
default $P$-even approach and allows the probing of the $P$-odd
contribution into the \CP asymmetry; the $P$-even contribution cancels
out. 

In the case 
of four-body meson decays, $P$-odd amplitudes can contribute only if the intermediate-resonance 
configuration is $VV$, $VT$ or $TT$ ($V$ and $T$ stand for vector and tensor meson respectively)
and if both resonances have helicities of either $\pm 1$ or $\pm
2$. For example, In the \decay{\Dz}{\pip\pim\pip\pim} decays there is one significant $P$-odd amplitude.
It is the one describing perpendicular helicity ($A_{\perp}$) of the $D^0\to\rhoz(770) \rhoz(770)$ decays.
Alternatively, in the partial wave basis, it is the amplitude corresponding to 
the $P$-wave $D^0\to\rhoz(770) \rhoz(770)$ decays, meaning relative orbital momentum 
of the two $\rho(770)$ mesons equal to 1. In such cases, the default
approach may be extended to make a complementary 
test of the $P$-odd \CP asymmetry. 

The complementarity of the $P$-even and $P$-odd \CP violation tests can be
illustrated in a simple simulation.  Simulated data samples, containing
one million \decay{\Dz}{\pip\pim\pip\pim} decays, are produced with MINT, a software package for amplitude analysis of
multi-body decays that has also been used by the CLEO
collaboration~\cite{Artuso:2012df}. The amplitude model used is based
on a preliminary version of that given in Ref.~\cite{dArgent:2016rbp}.  
\CP violation is introduced by changing the amplitude or phase of the
$P$-wave $D^0\to\rhoz(770) \rhoz(770)$ decays compared with the
\Dzb decays. Table~\ref{tab:p-odd_tests} shows that clear
sensitivity to the amplitude change is obtained in the $P$-even test
and to the phase change in the $P$-odd test.

\begin{table}[bhpt]
\centering  

\begin{tabular}{c|c|c}
\hline 
Asymmetries in & {$P$-even test }  & {$P$-odd test } \\
\decay{\Dz}{\rhoz\rhoz}($P$-wave) & $p$-value (fit) & $p$-value (fit) \\
\hline
$\Delta$phase 4$^\circ$, $\Delta$Amp 0 & $0.30^{+0.03 }_{ -0.03 }$ &  {\cellcolor{lightgray}{$1.95^{+0.06 }_{ -1.95 }\times10^{-4}$}} \\
\hline
$\Delta$phase 0$^\circ$, $\Delta$Amp 4 &
                                                       {\cellcolor{lightgray}{$3.02^{+ 1.2}_{ -0.9 }\times10^{-3}$}} & $0.41^{+0.03 }_{ -0.03 }$\\
\hline 
\end{tabular} 

 \caption{\small $p$-values for \decay{\Dz}{\pip\pim\pip\pim} simulation
   samples with phase and amplitude \CP asymmetries in
   $D^0\to\rhoz(770) \rhoz(770)$($P$-wave) (see text). Results from both
   $P$-even and $P$-odd \CP violation tests are given. The $p$-values
   are extracted from fits with a GEV function. 
   The cells shaded in gray demonstrate sensitivity to the simulated \CP
   violation scenarios.
} 
\label{tab:p-odd_tests}
\end{table}

\section{Implementation of unbinned techniques}
\label{sec:GPU}

The principal drawback of a number of unbinned statistical methods is
the computational time required for large sample sizes. In methods, such
as the energy test, that require the pairwise distance between all events in
the sample to be calculated, the computational time grows
quadratically with the sample size. Furthermore, in the energy test a
significant number of permutations are required for the random
comparison samples to get a sufficient precision on the probabilistic
interpretation of the $T$ value, for example to demonstrate that
evidence ($>3\sigma$) for \CP violation was observed over one thousand
permutations would be needed. In probing \CP violation in $b$-hadron decays the
computational constraints are typically not a limitation in these
methods currently. However, in multi-body decay channels of interest
in charm physics data samples of order one
million events are available at the \lhcb experiment.
At this sample size one permutation for the energy tests requires
around one day of CPU time on a typical computing node. Consequently the generation of 1000
permutations would require significant computational resources.
 
Even though modern
multi-core CPUs have realised thread-level parallelism,  the number of parallel threads is still
very limited. This has been overcome in our studies by implementation on Graphical
Processing Units (GPUs). A GPU  is  a  specialised  electronic  circuit  designed  to  rapidly
manipulate and alter memory to accelerate the creation of images for output to a display.  Their
highly parallel structure makes them more efficient than CPUs for algorithms where large blocks
of data are processed in parallel. Our implementation utilises the Compute
Unified Device Architecture (CUDA) and Thrust library developed by
NVIDIA~\cite{Thrust}. The parallelisation is utilised for the
calculation of the pairwise event distances.
One permutation for one million events takes approximately 30 minutes
to be computed on the two GPU
systems utilised (NVIDIA M2070, NVIDIA K40c). Both manual and Grid
submissions systems have been used. This implementation has made  the
energy test computationally feasible for the  first time in
\CP violation searches with large data sets. The code of this
implementation of the energy test, Manet, has been made available~\cite{EnergyTestSourceForge}.

\section{Conclusions}
\label{sec:conclusions}

The performance and optimisation of two techniques for model-independent searches for direct \CP violation in
multi-body decays is discussed. It is shown that binned
comparisons are best performed with a smaller number of bins than has
previously been used in some of the literature. The potential
advantages of unbinned techniques are discussed and demonstrated, and
an approach to account for the presence of background is suggested.
An implementation of the unbinned energy test technique is provided on
GPUs, which renders this method feasible for the largest data sample
sizes currently available at experiments.

Considerations in the choice, and symmetrisation, of the coordinates
used to describe the phase-space of three and four-body decays are
discussed, with specific examples given.
A novel method for analysing $P$-odd \CP violation in multi-body decays
in unbinned model independent searches is presented.

\section{Acknowledgements}

The authors wish to acknowledge the contributions to this work from
the following  MPhys and summer students at the University of
Manchester over the past three years: William Fawcett, Constantin
Weisser, Nicholas Bedford, Jezabel Boni, Samuel Dysch, Connor Graham.
We wish to thank Samuel Harnew, Jonas Rademacker, Gediminas Sarpis, and Mike
Williams for illuminating discussions. We thank Mike Sokoloff,
Andrew McNab and Sabah Salih for their assistance with computing. 
GPU facilities at the University of Manchester and the Ohio
Supercomputer centre have been utilised in this work.
This work was supported by the STFC grant numbers ST/J004332/1, ST/K003410/1, and ST/N000374/1.











\addcontentsline{toc}{section}{References}
\setboolean{inbibliography}{true}
\bibliographystyle{LHCb}
\bibliography{main,LHCb-PAPER,LHCb-CONF,LHCb-DP,LHCb-TDR}

\ifx\mcitethebibliography\mciteundefinedmacro
\PackageError{LHCb.bst}{mciteplus.sty has not been loaded}
{This bibstyle requires the use of the mciteplus package.}\fi
\providecommand{\href}[2]{#2}
\begin{mcitethebibliography}{10}
\mciteSetBstSublistMode{n}
\mciteSetBstMaxWidthForm{subitem}{\alph{mcitesubitemcount})}
\mciteSetBstSublistLabelBeginEnd{\mcitemaxwidthsubitemform\space}
{\relax}{\relax}

\bibitem{Kobayashi:1973fv}
M.~Kobayashi and T.~Maskawa, \ifthenelse{\boolean{articletitles}}{\emph{{\CP
  violation in the renormalizable theory of weak interaction}},
  }{}\href{http://dx.doi.org/10.1143/PTP.49.652}{Prog.\ Theor.\ Phys.\
  \textbf{49} (1973) 652}\relax
\mciteBstWouldAddEndPuncttrue
\mciteSetBstMidEndSepPunct{\mcitedefaultmidpunct}
{\mcitedefaultendpunct}{\mcitedefaultseppunct}\relax
\EndOfBibitem
\bibitem{OurNewPaper}
LHCb, R.~Aaij {\em et~al.}, \ifthenelse{\boolean{articletitles}}{\emph{{Search
  for $CP$ violation in the phase space of $D^0\rightarrow\pi^+\pi^-\pi^+\pi^-$
  decays}}, }{}\href{http://arxiv.org/abs/1612.03207}{{\tt
  arXiv:1612.03207}}\relax
\mciteBstWouldAddEndPuncttrue
\mciteSetBstMidEndSepPunct{\mcitedefaultmidpunct}
{\mcitedefaultendpunct}{\mcitedefaultseppunct}\relax
\EndOfBibitem
\bibitem{Dalitz:1953cp}
R.~H. Dalitz, \ifthenelse{\boolean{articletitles}}{\emph{{On the analysis of
  tau-meson data and the nature of the tau-meson}},
  }{}\href{http://dx.doi.org/10.1080/14786441008520365}{Phil.\ Mag.\
  \textbf{44} (1953) 1068}\relax
\mciteBstWouldAddEndPuncttrue
\mciteSetBstMidEndSepPunct{\mcitedefaultmidpunct}
{\mcitedefaultendpunct}{\mcitedefaultseppunct}\relax
\EndOfBibitem
\bibitem{Aubert:2008yd}
BaBar collaboration, B.~Aubert {\em et~al.},
  \ifthenelse{\boolean{articletitles}}{\emph{{Search for \CP Violation in
  Neutral D Meson Cabibbo-suppressed Three-body Decays}},
  }{}\href{http://dx.doi.org/10.1103/PhysRevD.78.051102}{Phys.\ Rev.\
  \textbf{D78} (2008) 051102}, \href{http://arxiv.org/abs/0802.4035}{{\tt
  arXiv:0802.4035}}\relax
\mciteBstWouldAddEndPuncttrue
\mciteSetBstMidEndSepPunct{\mcitedefaultmidpunct}
{\mcitedefaultendpunct}{\mcitedefaultseppunct}\relax
\EndOfBibitem
\bibitem{Aaij:2013jxa}
LHCb collaboration, R.~Aaij {\em et~al.},
  \ifthenelse{\boolean{articletitles}}{\emph{{Search for \CP violation in the
  decay $D^+ \to \pi^-\pi^+\pi^+$}},
  }{}\href{http://dx.doi.org/10.1016/j.physletb.2013.12.035}{Phys.\ Lett.\
  \textbf{B728} (2014) 585}, \href{http://arxiv.org/abs/1310.7953}{{\tt
  arXiv:1310.7953}}\relax
\mciteBstWouldAddEndPuncttrue
\mciteSetBstMidEndSepPunct{\mcitedefaultmidpunct}
{\mcitedefaultendpunct}{\mcitedefaultseppunct}\relax
\EndOfBibitem
\bibitem{LHCb-PAPER-2014-054}
LHCb collaboration, R.~Aaij {\em et~al.},
  \ifthenelse{\boolean{articletitles}}{\emph{{Search for $CP$ violation in
  $D^0\to\pi^-\pi^+\pi^0$ decays with the energy test}},
  }{}\href{http://dx.doi.org/10.1016/j.physletb.2014.11.043}{Phys.\ Lett.\
  \textbf{B740} (2015) 158}, \href{http://arxiv.org/abs/1410.4170}{{\tt
  arXiv:1410.4170}}\relax
\mciteBstWouldAddEndPuncttrue
\mciteSetBstMidEndSepPunct{\mcitedefaultmidpunct}
{\mcitedefaultendpunct}{\mcitedefaultseppunct}\relax
\EndOfBibitem
\bibitem{LHCb-PAPER-2013-057}
LHCb collaboration, R.~Aaij {\em et~al.},
  \ifthenelse{\boolean{articletitles}}{\emph{{Search for $CP$ violation in the
  decay $D^+ \to \pi^- \pi^+ \pi^+$}},
  }{}\href{http://dx.doi.org/10.1016/j.physletb.2013.12.035}{Phys.\ Lett.\
  \textbf{B728} (2014) 585}, \href{http://arxiv.org/abs/1310.7953}{{\tt
  arXiv:1310.7953}}\relax
\mciteBstWouldAddEndPuncttrue
\mciteSetBstMidEndSepPunct{\mcitedefaultmidpunct}
{\mcitedefaultendpunct}{\mcitedefaultseppunct}\relax
\EndOfBibitem
\bibitem{Link:2005th}
FOCUS collaboration, J.~M. Link {\em et~al.},
  \ifthenelse{\boolean{articletitles}}{\emph{{Search for T violation in charm
  meson decays}},
  }{}\href{http://dx.doi.org/10.1016/j.physletb.2005.07.024}{Phys.\ Lett.\
  \textbf{B622} (2005) 239}, \href{http://arxiv.org/abs/hep-ex/0506012}{{\tt
  arXiv:hep-ex/0506012}}\relax
\mciteBstWouldAddEndPuncttrue
\mciteSetBstMidEndSepPunct{\mcitedefaultmidpunct}
{\mcitedefaultendpunct}{\mcitedefaultseppunct}\relax
\EndOfBibitem
\bibitem{Link:2005ft}
FOCUS collaboration, J.~M. Link {\em et~al.},
  \ifthenelse{\boolean{articletitles}}{\emph{{Study of the decay asymmetry
  parameter and \CP violation parameter in the \decay{\Lc}{\Lambda\pip}
  decay}}, }{}\href{http://dx.doi.org/10.1016/j.physletb.2006.01.017}{Phys.\
  Lett.\  \textbf{B634} (2006) 165},
  \href{http://arxiv.org/abs/hep-ex/0509042}{{\tt arXiv:hep-ex/0509042}}\relax
\mciteBstWouldAddEndPuncttrue
\mciteSetBstMidEndSepPunct{\mcitedefaultmidpunct}
{\mcitedefaultendpunct}{\mcitedefaultseppunct}\relax
\EndOfBibitem
\bibitem{Bediaga:2009tr}
I.~Bediaga {\em et~al.}, \ifthenelse{\boolean{articletitles}}{\emph{{On a \CP
  anisotropy measurement in the Dalitz plot}},
  }{}\href{http://dx.doi.org/10.1103/PhysRevD.80.096006}{Phys.\ Rev.\
  \textbf{D80} (2009) 096006}, \href{http://arxiv.org/abs/0905.4233}{{\tt
  arXiv:0905.4233}}\relax
\mciteBstWouldAddEndPuncttrue
\mciteSetBstMidEndSepPunct{\mcitedefaultmidpunct}
{\mcitedefaultendpunct}{\mcitedefaultseppunct}\relax
\EndOfBibitem
\bibitem{Aaij:2011cw}
LHCb collaboration, R.~Aaij {\em et~al.},
  \ifthenelse{\boolean{articletitles}}{\emph{{Search for \CP violation in
  $D^{+} \to K^{-}K^{+}\pi^{+}$ decays}},
  }{}\href{http://dx.doi.org/10.1103/PhysRevD.84.112008}{Phys.\ Rev.\
  \textbf{D84} (2011) 112008}, \href{http://arxiv.org/abs/1110.3970}{{\tt
  arXiv:1110.3970}}\relax
\mciteBstWouldAddEndPuncttrue
\mciteSetBstMidEndSepPunct{\mcitedefaultmidpunct}
{\mcitedefaultendpunct}{\mcitedefaultseppunct}\relax
\EndOfBibitem
\bibitem{Laura}
T.~Latham, J.~Back, and P.~Harrison,
  \ifthenelse{\boolean{articletitles}}{\emph{Laura++ {Dalitz} plot fitting
  package}, }{} \url{https://laura.hepforge.org/}.
\newblock [Accessed 10-June-2014]\relax
\mciteBstWouldAddEndPuncttrue
\mciteSetBstMidEndSepPunct{\mcitedefaultmidpunct}
{\mcitedefaultendpunct}{\mcitedefaultseppunct}\relax
\EndOfBibitem
\bibitem{Williams:2011cd}
M.~Williams, \ifthenelse{\boolean{articletitles}}{\emph{{Observing \CP
  violation in many-body decays}},
  }{}\href{http://dx.doi.org/10.1103/PhysRevD.84.054015}{Phys.\ Rev.\
  \textbf{D84} (2011) 054015}, \href{http://arxiv.org/abs/1105.5338}{{\tt
  arXiv:1105.5338}}\relax
\mciteBstWouldAddEndPuncttrue
\mciteSetBstMidEndSepPunct{\mcitedefaultmidpunct}
{\mcitedefaultendpunct}{\mcitedefaultseppunct}\relax
\EndOfBibitem
\bibitem{Aaij:2013sfa}
LHCb collaboration, R.~Aaij {\em et~al.},
  \ifthenelse{\boolean{articletitles}}{\emph{{Measurement of \CP violation in
  the phase space of $B^{\pm} \to K^{\pm} \pi^{+} \pi^{-}$ and $B^{\pm} \to
  K^{\pm} K^{+} K^{-}$ decays}},
  }{}\href{http://dx.doi.org/10.1103/PhysRevLett.111.101801}{Phys.\ Rev.\
  Lett.\  \textbf{111} (2013) 101801},
  \href{http://arxiv.org/abs/1306.1246}{{\tt arXiv:1306.1246}}\relax
\mciteBstWouldAddEndPuncttrue
\mciteSetBstMidEndSepPunct{\mcitedefaultmidpunct}
{\mcitedefaultendpunct}{\mcitedefaultseppunct}\relax
\EndOfBibitem
\bibitem{Aaij:2013bla}
LHCb collaboration, R.~Aaij {\em et~al.},
  \ifthenelse{\boolean{articletitles}}{\emph{{Measurement of \CP violation in
  the phase space of $B^{\pm} \rightarrow K^{+} K^{-} \pi^{\pm}$ and $B^{\pm}
  \rightarrow \pi^{+} \pi^{-} \pi^{\pm}$ decays}},
  }{}\href{http://dx.doi.org/10.1103/PhysRevLett.112.011801}{Phys.\ Rev.\
  Lett.\  \textbf{112} (2014), no.~1 011801},
  \href{http://arxiv.org/abs/1310.4740}{{\tt arXiv:1310.4740}}\relax
\mciteBstWouldAddEndPuncttrue
\mciteSetBstMidEndSepPunct{\mcitedefaultmidpunct}
{\mcitedefaultendpunct}{\mcitedefaultseppunct}\relax
\EndOfBibitem
\bibitem{Porter:2008mc}
F.~C. Porter, \ifthenelse{\boolean{articletitles}}{\emph{Testing consistency of
  two histograms}, }{}\href{http://arxiv.org/abs/0804.0380}{{\tt
  arXiv:0804.0380}}\relax
\mciteBstWouldAddEndPuncttrue
\mciteSetBstMidEndSepPunct{\mcitedefaultmidpunct}
{\mcitedefaultendpunct}{\mcitedefaultseppunct}\relax
\EndOfBibitem
\bibitem{Fasano:1987}
G.~Fasano and A.~Franceschini, \ifthenelse{\boolean{articletitles}}{\emph{{A
  multidimenional version of the Kolmogorov-Smirnov test}},
  }{}\href{http://dx.doi.org/10.1093/mnras/225.1.155}{Mon.\ Not.\ R.\ astr.\
  Soc.\  \textbf{225} (1987) 155}\relax
\mciteBstWouldAddEndPuncttrue
\mciteSetBstMidEndSepPunct{\mcitedefaultmidpunct}
{\mcitedefaultendpunct}{\mcitedefaultseppunct}\relax
\EndOfBibitem
\bibitem{Williams:2010vh}
M.~Williams, \ifthenelse{\boolean{articletitles}}{\emph{{How good are your
  fits? Unbinned multivariate goodness-of-fit tests in high energy physics}},
  }{}\href{http://dx.doi.org/10.1088/1748-0221/5/09/P09004}{JINST \textbf{5}
  (2010) P09004}, \href{http://arxiv.org/abs/1006.3019}{{\tt
  arXiv:1006.3019}}\relax
\mciteBstWouldAddEndPuncttrue
\mciteSetBstMidEndSepPunct{\mcitedefaultmidpunct}
{\mcitedefaultendpunct}{\mcitedefaultseppunct}\relax
\EndOfBibitem
\bibitem{doi:10.1080/00949650410001661440}
B.~Aslan and G.~Zech, \ifthenelse{\boolean{articletitles}}{\emph{New test for
  the multivariate two-sample problem based on the concept of minimum energy},
  }{}\href{http://dx.doi.org/10.1080/00949650410001661440}{J.\ Stat.\ Comput.\
  Simul.\  \textbf{75} (2005) 109}\relax
\mciteBstWouldAddEndPuncttrue
\mciteSetBstMidEndSepPunct{\mcitedefaultmidpunct}
{\mcitedefaultendpunct}{\mcitedefaultseppunct}\relax
\EndOfBibitem
\bibitem{Aslan2005626}
B.~Aslan and G.~Zech, \ifthenelse{\boolean{articletitles}}{\emph{Statistical
  energy as a tool for binning-free, multivariate goodness-of -fit tests,
  two-sample comparison and unfolding},
  }{}\href{http://dx.doi.org/10.1016/j.nima.2004.08.071}{Nucl.\ Instrum.\
  Meth.\  \textbf{A537} (2005) 626 }\relax
\mciteBstWouldAddEndPuncttrue
\mciteSetBstMidEndSepPunct{\mcitedefaultmidpunct}
{\mcitedefaultendpunct}{\mcitedefaultseppunct}\relax
\EndOfBibitem
\bibitem{Rosenbaum}
P.~R. Rosenbaum, \ifthenelse{\boolean{articletitles}}{\emph{{An exact
  distribution-free test comparing two multivariate distributions based on
  adjacency}}, }{}\href{http://dx.doi.org/10.1111/j.1467-9868.2005.00513.x}{J.\
  R.\ Statist.\ Soc.\ B \textbf{67} (2005) 515}\relax
\mciteBstWouldAddEndPuncttrue
\mciteSetBstMidEndSepPunct{\mcitedefaultmidpunct}
{\mcitedefaultendpunct}{\mcitedefaultseppunct}\relax
\EndOfBibitem
\bibitem{LHCb-PAPER-2014-052}
LHCb collaboration, R.~Aaij {\em et~al.},
  \ifthenelse{\boolean{articletitles}}{\emph{{Search for the lepton flavour
  violating decay $\tau^-\to \mu^-\mu^+\mu^-$}},
  }{}\href{http://dx.doi.org/10.1007/JHEP10(2015)121}{JHEP \textbf{02} (2015)
  121}, \href{http://arxiv.org/abs/1409.8548}{{\tt arXiv:1409.8548}}\relax
\mciteBstWouldAddEndPuncttrue
\mciteSetBstMidEndSepPunct{\mcitedefaultmidpunct}
{\mcitedefaultendpunct}{\mcitedefaultseppunct}\relax
\EndOfBibitem
\bibitem{Byckling}
E.~Byckling and K.~Kajantie, {\em Particle Kinematics}, University of
  Jyvaskyla, Jyvaskyla, Finland, 1971\relax
\mciteBstWouldAddEndPuncttrue
\mciteSetBstMidEndSepPunct{\mcitedefaultmidpunct}
{\mcitedefaultendpunct}{\mcitedefaultseppunct}\relax
\EndOfBibitem
\bibitem{focus}
FOCUS collaboration, J.~M. Link {\em et~al.},
  \ifthenelse{\boolean{articletitles}}{\emph{{Study of the $D^0 \to \pi^{-}
  \pi^{+} \pi^{-} \pi^{+}$ decay}},
  }{}\href{http://dx.doi.org/10.1103/PhysRevD.75.052003}{Phys.\ Rev.\
  \textbf{D75} (2007) 052003}, \href{http://arxiv.org/abs/hep-ex/0701001}{{\tt
  arXiv:hep-ex/0701001}}\relax
\mciteBstWouldAddEndPuncttrue
\mciteSetBstMidEndSepPunct{\mcitedefaultmidpunct}
{\mcitedefaultendpunct}{\mcitedefaultseppunct}\relax
\EndOfBibitem
\bibitem{Artuso:2012df}
CLEO collaboration, M.~Artuso {\em et~al.},
  \ifthenelse{\boolean{articletitles}}{\emph{{Amplitude analysis of $D^0\to
  K^+K^-\pi^+\pi^-$}},
  }{}\href{http://dx.doi.org/10.1103/PhysRevD.85.122002}{Phys.\ Rev.\
  \textbf{D85} (2012) 122002}, \href{http://arxiv.org/abs/1201.5716}{{\tt
  arXiv:1201.5716}}\relax
\mciteBstWouldAddEndPuncttrue
\mciteSetBstMidEndSepPunct{\mcitedefaultmidpunct}
{\mcitedefaultendpunct}{\mcitedefaultseppunct}\relax
\EndOfBibitem
\bibitem{LHCb-PAPER-2014-046}
LHCb collaboration, R.~Aaij {\em et~al.},
  \ifthenelse{\boolean{articletitles}}{\emph{{Search for $CP$ violation using
  $T$-odd correlations in $D^0 \to K^+K^-\pi^+\pi^-$ decays}},
  }{}\href{http://dx.doi.org/10.1007/JHEP10(2014)005}{JHEP \textbf{10} (2014)
  005}, \href{http://arxiv.org/abs/1408.1299}{{\tt arXiv:1408.1299}}\relax
\mciteBstWouldAddEndPuncttrue
\mciteSetBstMidEndSepPunct{\mcitedefaultmidpunct}
{\mcitedefaultendpunct}{\mcitedefaultseppunct}\relax
\EndOfBibitem
\bibitem{Valencia}
G.~Valencia, \ifthenelse{\boolean{articletitles}}{\emph{Angular correlations in
  the decay {$B\to VV$} and {\CP} violation},
  }{}\href{http://dx.doi.org/10.1103/PhysRevD.39.3339}{Phys.\ Rev.\ D
  \textbf{39} (1989) 3339}\relax
\mciteBstWouldAddEndPuncttrue
\mciteSetBstMidEndSepPunct{\mcitedefaultmidpunct}
{\mcitedefaultendpunct}{\mcitedefaultseppunct}\relax
\EndOfBibitem
\bibitem{dArgent:2016rbp}
P.~d'Argent {\em et~al.}, \ifthenelse{\boolean{articletitles}}{\emph{{Amplitude
  analysis of $D^{0} \rightarrow \pi^{+} \pi^{-} \pi^{+} \pi^{-}$ decays using
  CLEO-c data}}, }{} in {\em {8th International Workshop on Charm Physics
  (Charm 2016) Bologna, Italy, September 5-9, 2016}}, 2016.
\newblock \href{http://arxiv.org/abs/1611.09253}{{\tt arXiv:1611.09253}}\relax
\mciteBstWouldAddEndPuncttrue
\mciteSetBstMidEndSepPunct{\mcitedefaultmidpunct}
{\mcitedefaultendpunct}{\mcitedefaultseppunct}\relax
\EndOfBibitem
\bibitem{Thrust}
{{NVIDIA Corporation,} {\it Thrust quick start guide},
  \href{http://docs.nvidia.com/cuda/pdf/Thrust_Quick_Start_Guide.pdf}{DU-06716-001
  (2014)}}\relax
\mciteBstWouldAddEndPuncttrue
\mciteSetBstMidEndSepPunct{\mcitedefaultmidpunct}
{\mcitedefaultendpunct}{\mcitedefaultseppunct}\relax
\EndOfBibitem
\bibitem{EnergyTestSourceForge}
J.~Brodzicka {\em et~al.}, \ifthenelse{\boolean{articletitles}}{\emph{Energy
  test implmentation on {GPUs}}, }{} \url{https://www.hepforge.org/}.
\newblock [Accessed 26-August-2016]\relax
\mciteBstWouldAddEndPuncttrue
\mciteSetBstMidEndSepPunct{\mcitedefaultmidpunct}
{\mcitedefaultendpunct}{\mcitedefaultseppunct}\relax
\EndOfBibitem
\end{mcitethebibliography}

\end{document}